\documentclass{aa0339}
\usepackage{graphicx}

\def\la{\;\raise0.3ex\hbox{$<$\kern-0.75em\raise-1.1ex\hbox{$\sim$}}\;}
\def\ga{\;\raise0.3ex\hbox{$>$\kern-0.75em\raise-1.1ex\hbox{$\sim$}}\;}

\begin{document}

\title{A catalogue of absorption-line systems in QSO spectra}
\author{
A.~I. Ryabinkov 
\and
A.~D. Kaminker 
\and
D.~A. Varshalovich 
}
\institute{
Ioffe Physical Technical Institute,
         Politekhnicheskaya 26, 194021 St.~Petersburg, Russia
\\
{\em  
 calisto@rbcmail.ru,
 kam@astro.ioffe.ru,
 varsh@astro.ioffe.ru
}}
\offprints{A.D.\ Kaminker}
 
 \date{}
 
\abstract
{
We present a new 
catalog of absorption-line systems 
identified in the quasar 
spectra. It contains data on 821 QSOs
and 8558 absorption systems comprising
16~139 absorption lines 
with measured redshifts
in the QSO spectra.
The catalog includes
absorption-line systems consisting of
lines of heavy elements, 
lines of neutral hydrogen,
Lyman limit systems,
damped Ly$\alpha$ absorption systems,
and broad absorption-line systems. 
The catalog is available in electronic form at 
the CDS via anonymous ftp to {\it cdsarc.u-strasbg.fr} (130.79.128.5)
or via http://{\it cdsweb.u-strasbg.fr/cgi-bin/qcat?J/A+A/412/707}
and at {\it www.ioffe.ru/astro/QC}. 
Using the data of the present catalog 
we also discuss     
redshift distributions of absorption-line systems.
}
      
\titlerunning{A catalogue of absorption-line systems}
\authorrunning{A.~I.\ Ryabinkov, A.~D.\ Kaminker, D.~A.\ Varshalovich}
\maketitle
 
\section{Introduction}
\label{sect-intro}
Absorption lines 
and absorption-line systems (ALSs)
observed in the spectra of QSOs  
contain fundamental information on 
the distribution of matter 
between the observer and the QSO,
and on physical processes  
in the Universe in different epochs 
of the cosmological evolution. 
To date, thousands of ALSs 
have been identified and  
their number grows persistently,
scattered over numerous 
sources.
This stimulates the creation of catalogs
of ALSs comprising the most complete data  
on the absorption lines and their systems. 
 
Catalogs of ALSs 
have been compiled many times.
We mention 
the early catalogs of Perry et al.\ (\cite{pbb78})
and Ellis \& Phillips\ (\cite{ep78}),
and the later vast QSO catalogs  
of Hewitt \& Burbidge 
(\cite{hb80},  \cite{hb87}, \cite{hb89}, and \cite{hb93})
which include also  
data on the ALSs 
detected in the QSO spectra.   
Junkkarinen et al.\ (\cite{jhb91}) and 
York et al.\ (\cite{yyccgm91})
created special ALS catalogs
most complete for that time.  
The new generation of telescopes (Keck, VLT, etc.)
has yielded a great amount of new spectroscopic data.
Some have been collected
in special catalogs including either
the results
of certain spectral investigations or  
the definite types of ALSs 
(e.g., Lyman limit systems -- LLS, 
damped Ly$\alpha$ absorption systems  -- DLA, 
broad absorption-line systems -- BAL, 
etc.). 
For example, the catalog by
Outram et al.\ (\cite{ossbclm01}) of the ALSs detected
in the 2dF QSO Redshift Survey or 
the catalogue of DLAs compiled by       
Curran et al.\ (\cite{cwmbcf02}). 
However, as far as we know, there are no
modern catalogs comprising comprehensive data on  
the ALSs registered to date. 
  
Our new catalog is an attempt to collect the basic
information on the ALSs 
in QSO spectra. The data are taken from
publications available up to January 2002.
The catalog
includes, in particular, all the  
data of the catalogs of
Junkkarinen et al.\ (\cite{jhb91}) and 
York et al.\ (\cite{yyccgm91}). 
The catalog consists of   
introduction (ReadMe),
Tables 1 and 2, and list of references,
which are available in electronic form  
at the CDS 
and at {\it www.ioffe.ru/astro/QC}.
 
\section{Description of the catalogue}
\label{sect-2}
{\bf Table 1} ({\it Quasars}) contains data on  
821 QSOs whose characteristics    
of spectral observations are collected in Table 2.
The QSO data are based on the 
catalog of Veron-Cetty \& Veron\ (\cite{vv01}).  
The following information on the QSOs is given:  
the QSO's name (J2000)  
identical to the name in Table 2 (see below)
and the name given 
in the catalog of Veron-Cetty \& Veron\ 
(\cite{vv01}); \,  
equatorial coordinates at 2000 and 1950, 
right ascensions $\alpha_{2000}$ and $\alpha_{1950}$,
and declinations  $\delta_{2000}$ and $\delta_{1950}$
in the order of increasing right ascensions; \,   
emission-line redshift $z_{\rm em}$; \, 
apparent magnitude $V$; \,  
absolute magnitude $M$. 
   
\noindent
{\bf Table 2} ({\it Absorption systems}) 
contains the following information 
on spectral observations of the QSOs 
and on the detected ALSs:  \\ 
(i) the name of QSOs  (J2000 and Q1950) 
which spectral observations 
are presented in literature,  \\
(ii)  {\it parameters of spectral observations} (see below), \\
(iii) {\it characteristics of absorption lines} 
combined in the systems. 
  
The {\it parameters of spectral observations} of 821 QSOs  
contain the following data: 
the interval of wavelengths (\AA)  
covered by the cited observations;
spectral resolution $R=\Delta \lambda_{\rm obs}$  (\AA);
signal-to-noise ratio $\langle S/N \rangle$, 
averaged over the entire spectral 
interval of observations; 
in some cases, the threshold (minimal) value 
of the absorption equivalent width in \AA $\,$ 
($W_{\rm min}$ or $W_{\rm rest,min}$
the observer or the rest reference 
frame)  used by the cited authors
as the criterion of line detection; 
the emission-line redshift $z_{\rm em}$;
references.  

Table~2  includes the
{\it characteristics of}  16~139 {\it absorp\-tion lines} 
detected in the spectra of 735 QSOs.
These lines form 8558 absorption systems.
In the spectra of 14 QSOs 
only Galactic interstellar lines
have been detected; 
characteristics of 
these lines are excluded. 
In the spectra of 72 QSOs 
the absorption lines
have not been detected or identified.
Table~2 comprises
3039 absorption-line systems 
including lines of heavy elements. 
For instance, we present 2871 resonance doublets 
of ions Mg II, Al III, C IV, Si IV, N V,
and O VI. The table comprises also
6063 systems containing 
lines of neutral hydrogen (HI).
In particular, there are 5554 systems
including only lines of HI,
146 LLSs, 195 DLA systems, 
and 39 BAL systems. 
LLSs are optically thick   
at the HI Lyman limit\  (912 \AA ). 
They correspond to HI column densities $N$(HI) 
$> 2\times 10^{17}~{\rm cm}^{-2}$.
DLA systems correspond to high HI 
column densities $N$(HI) 
$> 2\times 10^{20}~{\rm cm}^{-2}$.
BAL systems are characterized 
by wide absorption troughs produced by
ions with an outflow velocity 
extending  up to 60~000 km/s
relative to the QSO's. 
We rule out absorption systems consisting 
only of one heavy-element 
absorption line. 
  
The data are listed
in the order of increasing absorption-system redshifts. 
For heavy-element and HI systems 
we present the following information:
the absorption-system redshift $z_{\rm abs}$,
name of the ion identified,
laboratory wavelength $\lambda_{\rm lab}$ in \AA,
observed wavelength $\lambda_{\rm obs}$ in \AA, 
error $\sigma(\lambda_{\rm obs})$ 
of the value $\lambda_{\rm obs}$ at 
1$\sigma$ significance level,
absorption-line  equivalent width $W_{\rm obs}$~(\AA )
(measured in the observer frame) 
or column density of the ion $\log N$~(cm$^{-2}$),
error $\sigma(W_{\rm obs})$
(or $\sigma(\log N)$)  
of the values $W_{\rm obs}$   
or $\log N$~(cm$^{-2}$), respectively.    
For the  LLSs we give only 
the redshift $z_{\rm abs}$ of the absorption edge\  (912 \AA ). 
For the DLA systems we present: 
$z_{\rm abs}$,
the observational equivalent width $W_{\rm obs}$ (\AA) 
or HI~ column density  $\log N$(HI)~(cm$^{-2}$).  
For the BAL systems:  
the averaged value of $z_{\rm abs}$,
names of identified ions,
width of the trough $(\Delta z_{\rm abs})$.   

Table 2 contains also a few sets
of observational data on spectra of the QSOs
registered in different observations 
and presented by different authors.
Special signs indicate the cases where
absorption lines are blended by unidentified lines. 

All relevant notations and comments 
for users are given in the introduction (ReadMe).
The list of references contains  
literature sources 
quoted in Table~2.

The authors are planning to replenish     
the catalog regularly. 
Any remarks and comments  would be greatly 
appreciated. 

\section{Redshift distributions}
\label{sect-3} 

\begin{figure}[ht]
\setlength{\unitlength}{1mm}
\leavevmode
\includegraphics[width=86mm,  bb=80 205  500  465,clip]{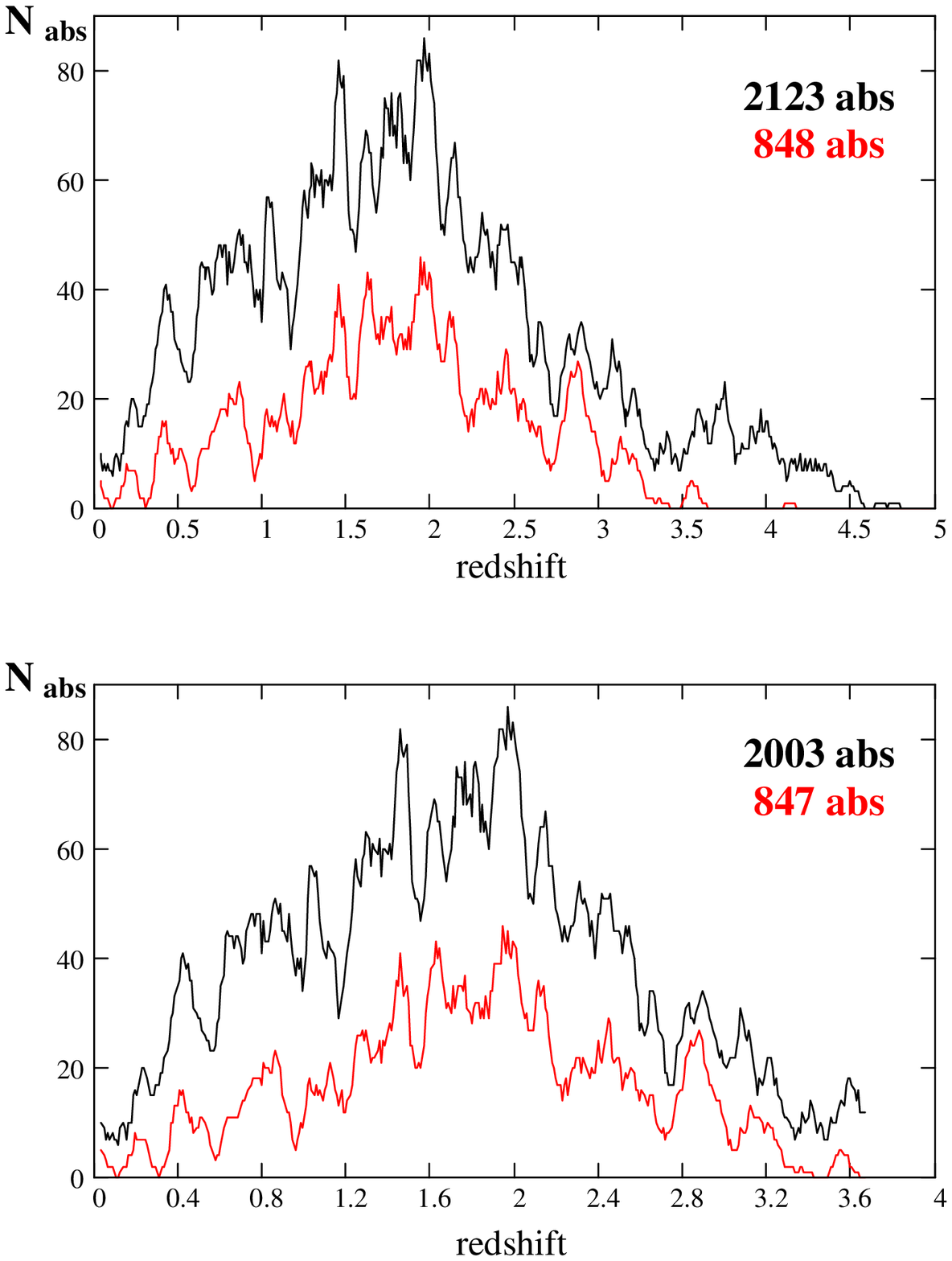}
\caption{
\footnotesize
Two $z$-distributions
of absorption systems observed in QSO spectra 
containing heavy-elements lines within 
the redshift interval $z=0$--3.7:
847 systems are      
obtained from the data 
of Junkkarinen et al.\  (\cite{jhb91}) 
and 2003 systems are     
obtained from the present catalogue (see text). 
}
\label{Metal91_03}   
\end{figure}

All absorption-line systems 
collected in the catalog belong to
the redshift interval from $z_{\rm min}=0.0033$
to $z_{\rm max}=4.93$. 
As an illustration, we 
compare (Fig.~\ref{Metal91_03}) two $z$-distributions
of the absorption systems including heavy-elements lines
within the redshift interval $z=0$--3.7.
One of them  
is obtained using the data of the catalog  
of Junkkarinen et al.\  (\cite{jhb91})  
and the other is based on 
the present catalog. 
In accordance with the data 
of Junkkarinen et al.\  (\cite{jhb91}) all
absorption redshifts  
fallen within an interval of 500~km~s$^{-1}$
are treated as a single system 
with a single $z_{\rm abs}$. 
We have compiled 2003 absorption systems
versus 847 systems from the catalog of
Junkkarinen et al.\  (\cite{jhb91}).  
Both distributions    
are obtained using the so-called 
sliding-average approach, in which  
a set of consecutive displacements of
the averaging  
bin $ \Delta_{\rm z} = 0.071$ 
is performed
along the $z$ axis with the step $\delta_{\rm z} = 0.01$.
 
As shown by
Ryabinkov et al.\ (\cite{rkv01}) 
from the data of
Junkkarinen et al.\  (\cite{jhb91})
the $z$-distribution of absorption-line systems 
displaces a pattern of alternating 
maxima (peaks) and minima (dips)
relative to a smooth curve. 
It is essential that their positions 
turn out to be independent 
of observation directions. 
Additionally,
the data revealed
a regularity (sort of periodicity) 
of the distribution with respect to 
some rescaled functions of $z$. 
This suggests that the derived distribution of
absorption matter is not only spatial but also
temporal in nature.
A comparison of the two 
$z$-distributions 
(Fig.~\ref{Metal91_03}) 
indicates
that the positions of all main peaks and dips
remain the same after the expansion of statistics
some of them are now more significant. 
 
These conclusions 
confirm the results of earlier statistical
analyses 
(Ryabinkov et al.\ \cite{rvk98} and 
Kaminker et al.\ \cite{krv00})
of $z$-distributions of  
C IV and Mg II 
absorption systems. 
Detailed statistical analysis
of such distributions
based on the present
catalogue will be done elsewhere.  

\begin{acknowledgements}
We are grateful to D.G. Yakovlev for useful remarks. 
The work was partly supported by RFBR grants No.\ 02-02-16278
and  03-07-90200. 

\end{acknowledgements}

\end{document}